# Two-Player Alternate Uses Test: A Controlled Testbed for Interactive Human–AI and Human–Human Co-Creation

Controlled, Interactive Human-AI Co-Creation

Babak Hemmatian[1], Anita Keshmirian[2], Yijun Lin, Shravan Ramamoorthy[3], Mostafa Jahadakbar[4], Elena Khuri-Reid[4], Jingyu Wang[5], Sina Hadjarab[6], Sigrid Veum[2], Pranav Gupta[4], Deepak Somaya[4], Lav Varshney[4]
[1]AI Innovation Institute, Stony Brook University [2]Forward College Berlin [3]Alinea [4]University of Illinois Urbana-Champaign [5]University of California San Diego [6]University of Chicago

Controlled research on AI ideation typically compares independent agents, while field studies of human–AI collaboration sacrifice experimental control. We introduce a controlled, two-player extension of the Alternate Uses Test (AUT) that enables comparison of human–human and human–AI co-creation under matched interactive conditions, alongside calibrated non-interactive baselines. The platform supports decomposition of performance into three typically confounded factors: participant traits, partner perceptions, and content dynamics. An in-person pilot (N = 62) demonstrates its utility. Under matched time limits, originality with a GPT-4 partner is statistically equivalent to that with a human partner. Approach motivation (BAS Drive) moderates whether interactive partnership benefits originality, and self-reported cognitive outsourcing predicts lower originality specifically in human–human dyads. Prior exposure to highly creative ideas improves later performance, suggesting a "seeding" intervention. We release the platform, code, and dataset as a shared testbed for controlled studies of human–AI co-creation.

## 1 INTRODUCTION

Large language models are deployed for creative tasks once thought exclusive to humans, and a psychology of AI creativity has emerged. Two research traditions dominate. Experimental work places humans and LLMs in parallel divergent-thinking tasks and compares their independent outputs (e.g., [14, 26]), yielding clean causal identification but no purchase on what happens during co-creation. Field studies of human–AI teaming in contexts like classrooms and workplaces (e.g., [3, 6]) preserve ecological validity but sacrifice experimental control over the interaction. The middle ground, controlled study of real-time co-creation between a human and an AI compared with human-human interaction, is less studied.

A handful of recent studies begin to close this gap. Doshi and Hauser [8] show that one-shot exposure to AI-generated story ideas raises individual originality while narrowing collective diversity, without conversation between participant and AI. Ashkinaze et al. [1] scale this passive-exposure paradigm to a large dynamic experiment. Shaer et al. [23] embed an LLM in asynchronous group brainwriting and validate LLM-based idea evaluation. Closest in design, Maier et al. [17] randomize interaction structure in one-on-one human–AI chat. Each isolates a different slice of the problem, but none preserves a classical divergent-thinking task across matched human–human, human–AI, and non-interactive conditions in a single apparatus so that interactivity, partner identity, conversation structure, and content exposure can be disentangled.

This gap connects to a long-running C&C concern with creativity as a situated, dynamically unfolding interactive process. Prior C&C work has quantified interaction dynamics in open-ended co-creation [5] and articulated design principles for socially interactive systems [16]; more recent work frames turn-taking, contribution type, communication, and feedback as the dimensions that shape human–AI co-creative interaction [20], with interaction modality and communication design influencing perceptions of the co-creative process [7, 19]. There is need to translate these interactional concerns into a controlled testbed where partner identity, interaction structure, perception, and content can be simultaneously tested as competing causes of co-creative outcomes, speaking to the C&C 2026 theme of "Creativity for Change."

We introduce a platform designed to address this gap, built around a two-player extension of the Alternate Uses Test (AUT). The AUT asks participants to generate unusual uses for an everyday object: for example, using a brick as a "doorstop" is common, while employing it as a "passive thermal battery" is original. Since Guilford [12], the AUT has been a leading measure of divergent thinking and, more recently, the standard task against which LLM creativity is benchmarked. In our platform, two players (two humans or a human and a conversational LLM) generate uses for a target object together in a shared chat interface under a fixed time limit and matched instructions. A third, non-interactive condition replaces the live partner with pre-rated sets of highly creative, uncreative, or GPT-4-only ideas, so that interaction effects can be estimated against a calibrated baseline.

### 1.1 Contributions

We contribute:

(1) a two-player, time-limited AUT chatroom supporting human–human, human–AI, and non-interactive baselines under closely matched procedures, with configurable LLM backend and interaction parameters;

(2) a pilot-tested dataset of 1,928 ideas from 62 participants with 6× redundant originality ratings; and

(3) pilot analyses illustrating the testbed's ability to decompose co-creative performance into three classes of factors typically confounded in prior work: 1) participant traits (e.g., socioemotional sensitivity, motivation), 2) partner effects (e.g., trait impact on partnerships with a human versus an LLM) and, 3) interaction dynamics and content exposure (e.g., prior idea exposure). We provide preliminary evidence from the pilot for several hypotheses enabled by the decomposition that are consistent with cognitive outsourcing and content-based "seeding" effects.

## 2 THEORETICAL BACKGROUND

We motivate our measure selection through a framework that distinguishes collaborative engagement from cognitive outsourcing. Intelligent behavior often reflects effective reliance on others' knowledge rather than individual reasoning [21, 24]. Hemmatian and Sloman [13] argue that a combination of sensitivity to social signals and motivational factors determine whether co-creation partners outsource (treating the partner as a passive resource) or collaborate (share intentionality and engage in iterative co-construction), with consequences for co-creation quality [22]. We use the Reading the Mind in the Eyes Test [2] as a canonical test of subtle social-cue processing as a component of collective intelligence [27], including in fully text-based interactions [9] and the Behavioral Inhibition/Activation Scale [4] for motivational factors.

Co-creation outcomes reflect not only agent traits but also perceptions of partner attributes that, accurate or not, dictate interaction patterns. To explore the role of partner perception, we measured two foundational dimensions of social perception, warmth and competence, which robustly govern how people evaluate and interact with social partners [10, 11]. We used responsiveness as a proxy for the warmth factor, since some of its original connotations in the social cognition literature do not neatly apply to interactions with conversational LLMs.

## 3 PLATFORM AND STUDY DESIGN

### 3.1 Platform Design

The platform is a browser-hosted, two-player chat application using free-tier authentication, database, and hosting. Environment variables control LLM backend, prompts, condition assignment, task items, chat length, and response wait times, so the platform generalizes beyond the pilot reported here. Source code and a live demo are available.

The interface follows standard messaging conventions (chat pane, input field, avatars) and is held constant across conditions to avoid confounding within-subject comparisons. While this design choice supports internal validity, we note that fixed interface features may influence absolute originality levels (see Limitations).

A key design constraint is interaction parity across partner types. In the human–AI condition, the OpenAI API enforces strict turn-taking (the model cannot access a user message while generating), and this constraint is made explicit in the instructions. Human–human interactions use the same interface with naturally occurring turn-taking, ensuring that all other interaction elements are matched as closely as possible.

Finally, participants are explicitly informed of partner identity (human or AI). This is both a study objective (since partner perception is itself a variable of interest) and a practical necessity, as latency and model verbosity made identity difficult to conceal in pre-experimental tests.

**AI configuration.** The GPT partner used gpt-4 via OpenAI's chat-completions API with temperature = 0.7, max_tokens = 3000, and server-enforced strict turn-taking. Human and AI partners received instructions matched on all task-defining elements (object, creative-use goal, chat-based turn-taking, rule against conventional uses). The AI's prompt added behavioral guardrails identified in piloting to neutralize model-side asymmetries (verbosity, meta-commentary): one idea per turn, omit boilerplate preambles, single-sentence responses. Instruction differences were kept minimal for effective cross-condition evaluation. A token-overlap filter rejected candidate responses overlapping >50% with any prior message. The full prompt and filter logic are pinned at commit edc61a3 in the public repository.

### 3.2 Experimental Design

Sixty-two University of Illinois Urbana-Champaign students signed up for in-person sessions in exchange for course credit (19 male, 42 female, 1 non-binary). Each participant completed three co-creation sessions within-subject in randomized order: human–human (HUM), human–GPT-4 (GPT), and non-interactive (CON for "constant"). Sessions used different target objects (book, fork, tin can; counterbalanced). After each 4-minute chat, participants curated the "best" creative uses inspired by the chat in a separate survey tab.

Half of the participants completed a 2-minute solo AUT before the sessions to serve as a personal baseline; the other half completed it after to control for practice effects.

**Non-interactive baselines.** In the CON condition, participants were shown one of three pre-rated sets of 16 ideas: highly creative human ideas, uncreative human ideas, or GPT-4-only ideas. The non-GPT sets were drawn from Stevenson et al. [25], filtered so two independent researchers agreed on the extremes of the originality scale. As a manipulation check, crowdsourced originality of CON-session ideas confirmed the intended quality gradient (low_con < gpt_con < high_con; $\beta_{low-gpt} = -0.48$, $\beta_{high-gpt} = +0.23$, both $p < .001$).

**Measures.** A small, a priori set of five partner perception questions followed the third chat, thus placed to avoid impacting interaction styles. Of the five questions, perceived outsourcing, responsiveness and competence are analyzed here because they index the theoretically motivated outsourcing as well as warmth/competence axes (§2).

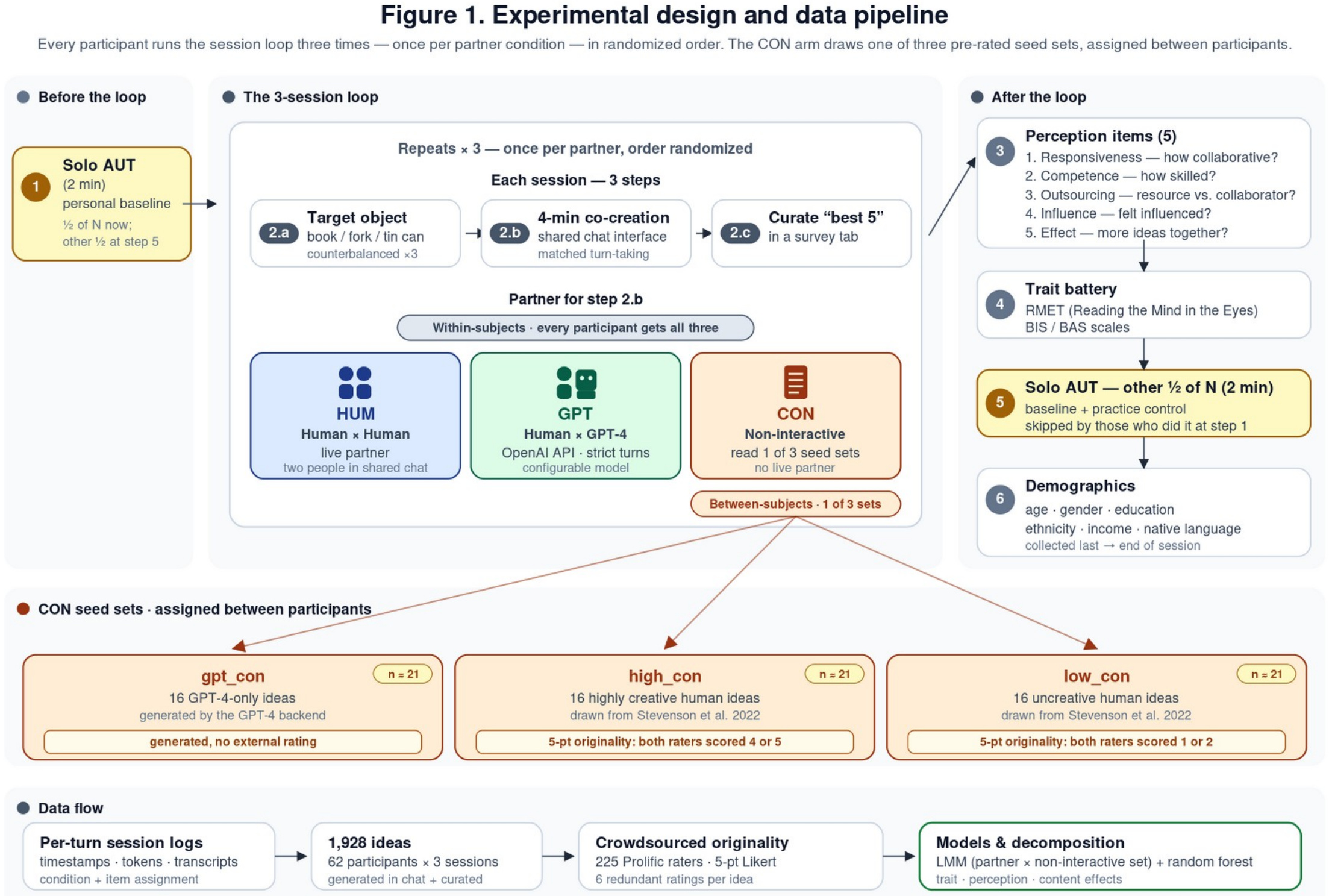

Figure 1: Experimental design and data pipeline. Each participant (N = 62) runs three randomized 4-minute co-creation sessions—once

with HUM (human partner), GPT (GPT-4 partner), and CON (one of three between-subjects pre-rated seed sets), bracketed by a 2-minute solo Alternate Uses Test as a personal baseline. After chat session 3, participants complete five perception items, the Reading the Mind in the Eyes (RMET) and Behavioral Inhibition/Activation Scale (BIS/BAS), and demographics. Curated ideas are rated by crowdsourced workers and modeled with a Linear Mixed-effects Model (LMM; partner × set) plus a random forest on traits. ***Text alternative****: Five-panel workflow diagram. Top-left panel ("Before the loop") shows step 1, a 2-minute solo AUT baseline completed by half the participants before the session loop. Top-center panel ("The 3-session loop") shows three steps repeated three times per participant (target object → 4-minute shared-chat co-creation → curating five best ideas), and three within-subjects partner conditions: HUM (human–human), GPT (human–GPT-4 with strict turn-taking), and CON (non-interactive, reading one of three seed sets). Top-right panel ("After the loop") lists post-session components: five perception items (responsiveness, competence, outsourcing, influence, effect), the RMET and BIS/BAS trait battery, the other half's solo AUT, and demographics. Middle row shows three between-subjects CON seed sub-arms (gpt_con: 16 GPT-4-only ideas; high_con: 16 highly creative human ideas from Stevenson et al.; low_con: 16 uncreative human ideas), with ~21 participants each. Bottom row shows the data pipeline: per-turn session logs → 1,928 ideas → crowdsourced originality ratings (225 Prolific raters, six per idea) → mixed-effects models and random-forest decomposition into trait, perception, and content effects.*

### 3.3 Psychological Measures

We used the Reading the Mind in the Eyes Test (RMET; [2]) as a behavioral index of socioemotional sensitivity. Recent work argues that RMET is better characterized as emotion recognition than Theory of Mind (ToM) [18]. Under any live interpretation, RMET indexes a trait-level tendency to attend to partners' states, which plausibly predicts textual co-creation through affect-reading, sustained partner-directed attention, or related processes. We chose RMET over self-report perspective-taking to avoid demand characteristics, and over longer behavioral ToM tasks to avoid verbal-ability confounds. We also administered the BIS/BAS scale [4] for examining motivational traits, reporting all subscales.

### 3.4 Originality Ratings

Following the classic AUT literature, we define originality as how unlikely others would be to generate the idea and utility as its practical usefulness [12, 25]. We report originality here; utility results are available on request.

**Rating protocol.** The dataset contains 1,923 distinct ideas rated by 226 crowdsourced workers (median 6 ratings per idea, range 5–12, with ~3% of ideas double-rated for reliability checks). Each idea received independent originality ratings on a 5-point Likert scale. Rater identity was modeled as a crossed random intercept in all mixed-effects analyses, absorbing severity/leniency differences (Var_rater = 0.29, SD = 0.54). Aggregate crowdsourced originality on the curated non-interactive sets was reasonably aligned with the two researchers' curation ratings ($r = .54$); a follow-up with validating research-assistant ratings is planned to triangulate.

## 4 ANALYSIS AND RESULTS

We illustrate how the platform separates baseline performance differences from trait-, perception-, and interaction-level correlates of co-creation. We fit a Linear Mixed-effects Model (LMM) with fixed effects of Partner type, non-interactive assignment, and their interaction, plus random intercepts for subject, item-within-subject, and rater. A random forest provides a nonlinear complement. Effect estimates are exploratory given the modest N, but limited power biases against finding effects: significant effects therefore survive a conservative test, and covariate results across model families are mutually corroborating.

### 4.1 Baseline Originality Across Partner Types

In the non-interactive condition, GPT-4 ideas were rated more original than the uncreative human set but less than the highly creative human set. Interactive HUM and GPT sessions yielded intermediate, statistically equivalent originality: $\beta_{GPT-HUM} = -0.11$, SE = 0.13, 95% CI [−0.37, +0.16], $p = .42$. TOST [15] bounded the maximum credible effect at $|\beta| \leq 0.33$, and a Bayesian re-estimation with a weakly informative Normal(0, 1) prior yielded substantial evidence in favor of the null ($BF_{01} = 10.95$; posterior median +0.06, 95% credible interval [−0.06, +0.20]). Under matched time limits, originality with a human partner and with a GPT-4 partner is statistically equivalent.

### 4.2 Determinants of Interactive Co-Creativity

We next ask which participant traits and partner perceptions predict originality under interactive conditions.

**Participant trait predictors.** Random-forest variable importance identified BAS subscales and RMET as the leading trait predictors of originality (Figure 2). Parametric tests revealed the signal comes from interactions, not main effects:

BAS Drive moderated the effect of partner type, with high-Drive participants extracting greater originality gains from interactive partners than from passive seed exposure (β_{BAS-Drive × GPT-vs-CON} = +0.20, 95% CI [+0.05, +0.38], $p = .034$, $BF_{10} = 3.37$). The same direction is replicated on top-quality rate ($p = .033$) and failure rate ($p = .030$). Disaggregated by partner (Figure 3), low-Drive participants are hurt mainly by pairing with fellow humans and high-Drive participants benefit mainly from GPT interaction. RMET was also positively associated with the within-subject range of originality ($r = +0.27$, $p = .037$), suggesting that socioemotional sensitivity may influence the spread of creative output (i.e., the willingness to take varied creative risks) rather than overall quality. The outsized impact of RMET on the random-forest results in Figure 2 suggests that socioemotional sensitivity modulated the variance of originality rather than shifting group means, meaning a linear aggregate model would have missed its effect. RMET and BAS Total main effects were equivalent to zero by Bayesian criteria ($BF_{01}$ = 19 and 26), consistent with condition-specific pathways for these predictors.

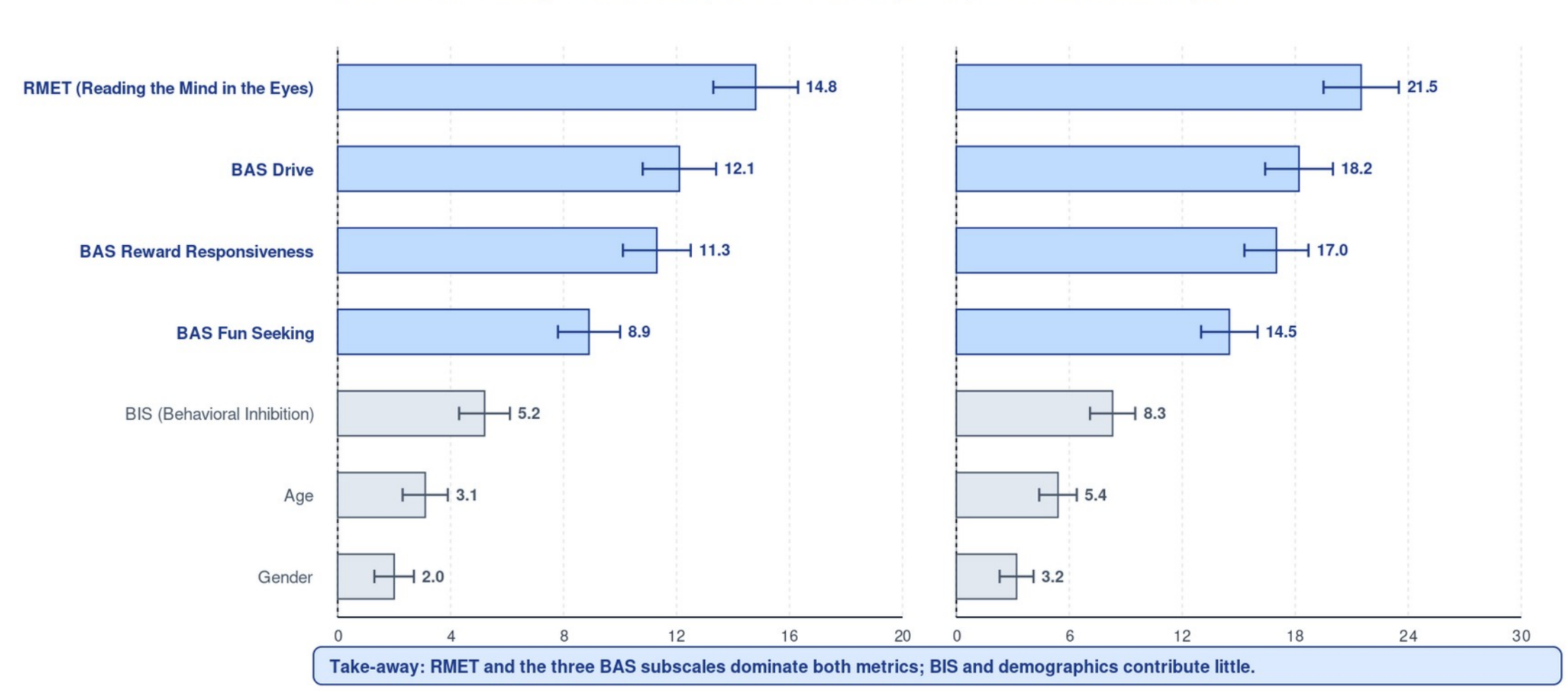

Figure 2: Random-forest variable importance for predicting originality, over 200 bootstrap replicates. Bars show mean importance; whiskers show ±1 SD across replicates. %IncMSE quantifies the increase in prediction error if the predictor were removed; IncNodePurity quantifies how much the predictor reduces error across tree splits. Reading the Mind in the Eyes Test and the three Behavioral Activation subscales dominate both metrics; Behavioral Inhibition Scale and demographics contribute little. Text alternative: horizontal bar charts showing variable importance on each metric, with RMET and BAS subscales highest.

**Partnership style predictor.** Self-reported outsourcing predicts lower originality when paired with humans ($r = -0.504$, $p = .024$, $n = 20$), but trends in the opposite direction when partnering with GPT-4 ($\beta = +0.19$, anecdotal in Bayesian terms). The direction reversal ($\Delta\beta \approx +0.35$) motivates a confirmatory test of an Outsourcing × Partner-type interaction in a larger sample, currently in progress with direct self- and partner-effort items. This outsourcing effect was not driven by perceived partner competence or responsiveness, neither of which robustly predicted originality at the proper between-subjects level ($BF_{01} = 16.40$ in favor of the null): the human-specific outsourcing reflects disengagement, not perceived partner quality.

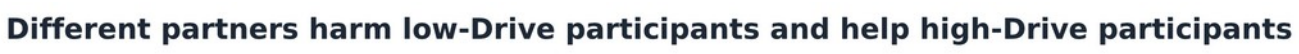

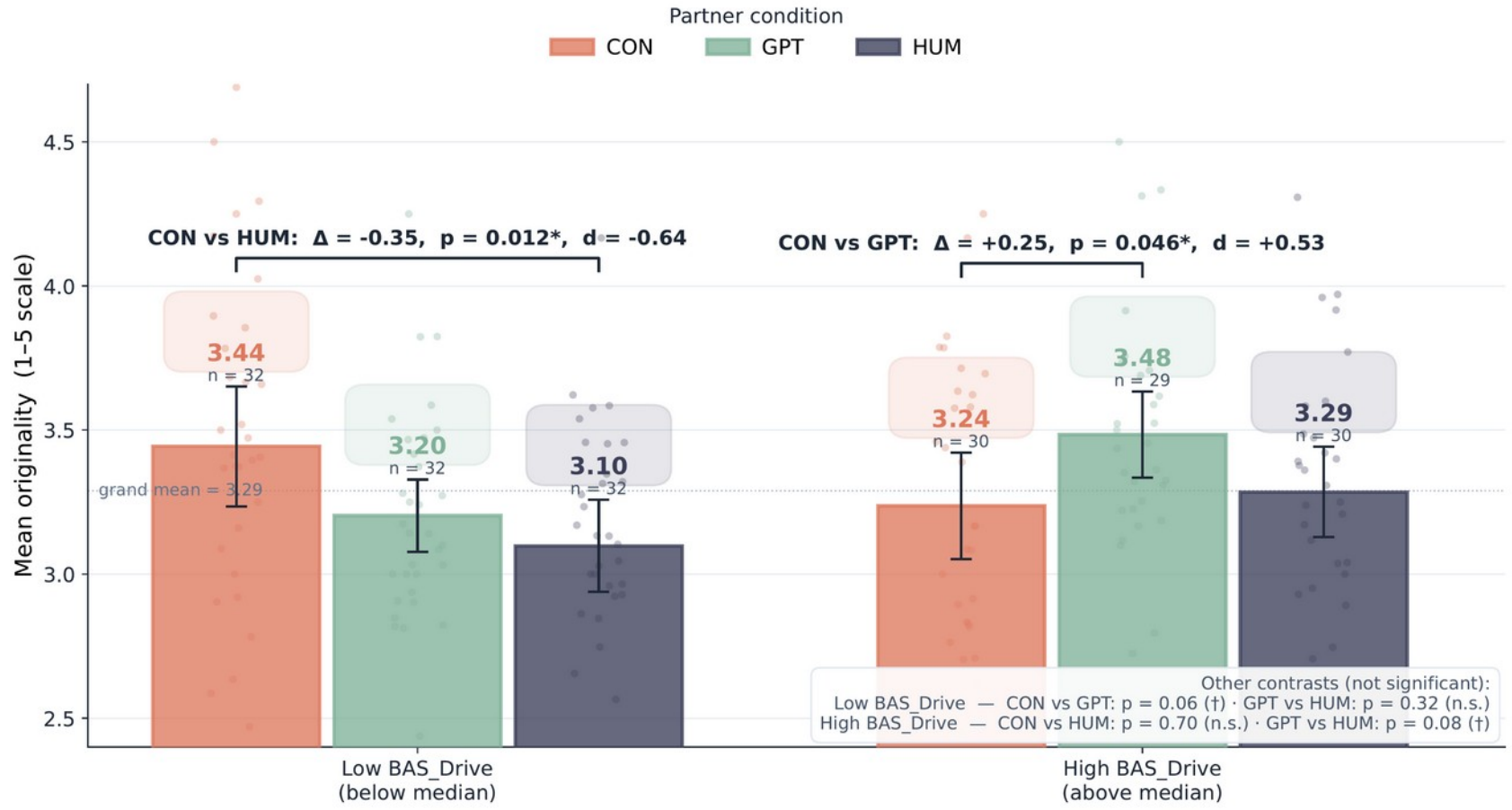

Figure 3: BAS Drive moderates whether interactive partnership helps creativity. Bars show mean originality (1–5 scale) per BAS Drive group × partner condition, with 95% CIs. Significant pairwise contrasts within each BAS Drive group ($p < .05$) are shown with brackets. Low-Drive participants are hurt by human partners (CON vs HUM $\Delta = -0.35$, $p = .012$, $d = -0.64$); high-Drive participants are helped by GPT partners (CON vs GPT $\Delta = +0.25$, $p = .046$, $d = +0.53$). Text alternative: bar chart with six bars, three partner conditions × two BAS Drive groups, showing the interaction described above.

### 4.3 Content Effects

Finally, we examine interaction-level effects. For participants whose first session was non-interactive, subsequent originality in their HUM and GPT sessions depended on the initial seed set ($F(2, 1963) = 4.68$, $p = .009$): prior exposure to highly creative human ideas was associated with higher subsequent originality with both human and AI partners ($\beta_{\text{high vs gpt}} = +0.32$, $p = .016$). This pattern points to a tractable “seeding” intervention: pre-exposing participants to highly original exemplars for a different target item before ideation can enhance later co-creation regardless of interactive partner type. Because CON sub-arms were unevenly assigned, we report this as an exploratory but directionally consistent finding that we are currently replicating in a larger sample.

## 5 DISCUSSION

Our results illustrate how a controlled interactive testbed can reveal patterns that are difficult to isolate in prior work, where personal traits, partner type, interaction structure and content are often confounded. Using this decomposition, robust patterns emerged despite the small sample size of our pilot study.

Under matched time limits, originality with a GPT-4 partner was statistically equivalent to originality with a human partner, as confirmed by TOST and Bayesian tests. This performance parity contrasts with prior reports of AI surpassing humans on independent divergent-thinking tasks [14, 26], illustrating that conclusions about AI versus human creativity depend critically on whether interaction is present and how it is structured.

Individual trait differences modulated whether interactive partnership benefited creativity. Approach motivation (BAS Drive) emerged as the strongest such moderator: low-Drive participants were consistently hurt by human partners, whereas high-Drive participants were reliably helped by GPT partners. That self-reported cognitive outsourcing predicted lower originality specifically in human–human dyads may explain the effect for low-Drive participants: The lower their trait motivation, the more likely participants were to outsource effort, resulting in worse outcomes. The forced turn-taking in the GPT condition may have partly offset this dynamic, though not enough for low-Drive participants to match their high-Drive peers. Together these findings are consistent with the distinction between collaborative engagement and cognitive outsourcing [13]. They also extend the COFI framework’s claim that interaction structure, particularly turn-taking and contribution type, determines whether an AI functions as tool, partner, or collaborator [20]: who benefits from a structure depends on the user’s motivational profile, not only the structure itself.

Sensitivity to one’s partner mattered as well. The Reading the Mind in the Eyes Test, which indexes attunement to others' mental states, was a top predictor of originality in a task with no visual cues. This suggests that a general orientation toward attending to partners helps co-creation. This is consistent with Woolley and colleagues' finding that social sensitivity predicts collective performance [27], including in fully text-mediated settings [9]. Whereas prior C&C work has emphasized how interaction modality and communication design shape co-creative perceptions [7, 19], our results point to a complementary, perceiver-side determinant: stable individual differences in attunement to partners can predict creative outcomes even when modality and communication design are held constant.

Beyond traits and perceptions, content exposure also mattered. Participants exposed to highly creative human ideas produced more original outputs in subsequent sessions with both human and GPT partners. Unlike trait-level predictors, seed content is directly manipulable, suggesting a concrete intervention: seed collaborators with highly original exemplars before ideation. This builds on the C&C tradition of quantifying co-creative interaction dynamics [5] by showing that the content a collaborator has seen, not only how they take turns, leaves a durable trace on what they produce next. It is also a tractable lever against the homogenization observed under passive AI exposure [8].

Given the modest pilot sample, these findings should be interpreted as exploratory and hypothesis-generating. However, the consistency of patterns across multiple analytical approaches, despite stringent design and statistical controls, suggests they reflect underlying signal rather than model-specific artifacts. The primary role of the pilot is to demonstrate that the platform can separately detect and relate trait-, perception-, and interaction-level influences within a single experimental framework. Future studies with similar design can confirm and expand on these initial findings.

## 6 LIMITATIONS

Partner identity is not blinded, since perceptual effects are themselves a study goal; this leaves AI-ness confounded with perception, but a blinded follow-up equalizing turn-taking and verbosity is feasible. Crowdsourced originality ratings are mitigated by redundancy and rater random effects, with an RA-rated replication planned. The interface and avatars were held constant for internal validity but may influence absolute originality, inviting a dedicated study. Human and AI instructions differ in behavioral guardrails needed for conversational parity, with all task-defining elements matched; the guardrails' influence invites future work. Finally, the scope is a single 4-minute AUT with English-speaking student participants; generalization across tasks, populations, and durations is an empirical question the platform supports.

## 7 CONCLUSION

We contribute a controlled, two-player AUT platform for the interactive study of human–human and human–AI co-creation, a pilot dataset of 1,923 rated ideas, and three exploratory findings: (1) originality in collaboration with a GPT-4 partner is statistically equivalent to that with a human partner under matched time limits; (2) socioemotional sensitivity and approach motivation are trait predictors of co-creation success, while self-reported cognitive outsourcing predicts lower originality specifically in human–human dyads; (3) prior exposure to highly creative ideas raises subsequent originality, suggesting a "seeding" intervention. We release the platform as a shared, pre-registerable testbed for studying the mechanisms of human–AI co-creation.